\documentclass[conference]{IEEEtran}
\usepackage{xspace}
\usepackage{xcolor}
\usepackage{amssymb}
\usepackage{amsmath}
\usepackage{graphicx}
\usepackage{url}
\usepackage{colortbl}  
\usepackage{siunitx}
\usepackage{enumitem}

\newcommand{\mymethod}{GNNVault\xspace}

\newif\ifcomment
\newcommand{\rd}[1]{\ifcomment {\color{purple}
\emph{[Ruyi: #1]}} \fi}
\newcommand{\ad}[1]{\ifcomment{\color{green}
\emph{[Adam: #1]}}\fi}
\newcommand{\yf}[1]{\ifcomment{\color{red}
\emph{[Fei: #1]}}\fi}

\begin{document}
\title{Graph in the Vault: Protecting Edge GNN Inference with Trusted Execution Environment}

\author{
  \IEEEauthorblockN{Ruyi Ding*, Tianhong Xu*, Aidong Adam Ding, Yunsi Fei}
  \IEEEauthorblockA{Northeastern University\\
  \{ding.ruy, xu.tianh, a.ding, y.fei\}@northeastern.edu}
}

\maketitle

\IEEEpeerreviewmaketitle

\begin{abstract}
Wide deployment of machine learning models on edge devices has rendered the model intellectual property (IP) and data privacy vulnerable. 
We propose \mymethod, the first secure Graph Neural Network (GNN) deployment strategy based on Trusted Execution Environment (TEE). 
\mymethod follows the design of  ``partition-before-training" and includes a private GNN rectifier to complement with a public backbone model. 
This way, both critical GNN model parameters and the private graph used during inference are protected within secure TEE compartments.
Real-world implementations with Intel SGX demonstrate that \mymethod safeguards GNN inference against state-of-the-art link stealing attacks with a negligible accuracy degradation ($<2$\%).
\end{abstract}

\section{Introduction} \label{sec: introduction}

On-device machine learning has emerged as an important paradigm for tasks requiring low latency and high privacy~\cite{dhar2021survey}. 
Applications like Apple's CoreML~\cite{coreml} and Huawei's MindSpore~\cite{mindspore} enable user personalization on local devices. 
This trend has also extended to Graph Neural Networks (GNNs)~\cite{zeng2022gnn, shao2021branchy}, ensuring the privacy of user data during inference for tasks such as community detection~\cite{shchur2019overlapping}, e-commerce personalization~\cite{zhu2019aligraph}, and recommender systems~\cite{wu2022graph}. 
However, local GNN inference grants users significant privileges to local models and data, introducing additional security vulnerabilities~\cite{xue2021dnn}.

There are two primary security threats to inference on local devices: model intellectual property (IP) theft and data privacy breach. 
Valuable model IP includes the hyper-parameters, trained parameters, and model's functionality~\cite{gongye2024side}. 
Data privacy issues involve attacks such as membership inference~\cite{shokri2017membership} and link stealing attack~\cite{he2021stealing}, which can exploit the white-box accessibility of local models.
Existing defenses, including  watermarking \cite{guo2018watermarking}, non-transferable learning \cite{wang2021non}, and homomorphic encryption~\cite{xie2019bayhenn}, are often passive, inaccurate, or computation-expensive. 
Recently, studies have begun leveraging Trusted Execution Environments (TEEs) to protect local models and data~\cite{zhang2024no, liu2023mirrornet}, by partitioning them into slices and deploying critical components within secure compartments.

However, current TEE-based secure deployment mainly targets datasets for tasks such as computer vision and large language models, which is not applicable to GNN inference. 
Specifically, GNN operations require updating the graph node representations (i.e., embeddings) based on information from their neighboring data points. 
Thus, \textit{the entire graph data, including node features and their connectivity (adjacency matrix), has to be stored locally during inference}, posing additional challenges for data privacy. 
Moreover, TEEs have limited memory capacities, while large graphs require significant storage for node features. 
To bridge this gap, we propose a new method for local GNN deployment, \mymethod, which effectively preserves both critical model parameters and the most important structural information (i.e., edges) using TEE.

\mymethod employs the ``partition-before-training'' strategy~\cite{zhang2024no}, where the GNN model is designed with a public backbone and a private GNN rectifier. Specifically, the public backbone is computation-intensive but inaccurate, which is trained with a substitute graph and deployed in the untrusted environment. 
The GNN rectifier leverages the real private graph structure to rectify the backbone's node embeddings. It is designed to be much smaller and resides within the TEE. 
Our experiments show that \mymethod achieves performance comparable to the original unprotected GNN while preventing information leakage of private parameters and edges. 
We also evaluate our design on a real system using Intel SGX, demonstrating the low overhead of \mymethod. 

\textbf{Our Contributions:} We propose \mymethod, the first secure deployment framework designed specifically for graph neural networks. Our work makes the following contributions:

\begin{enumerate}[leftmargin=*]
\item We propose various communication schemes between public backbones and private rectifiers in \mymethod, and evaluate the performance and implementation cost of rectifiers.
\item Software evaluation demonstrates that \mymethod achieves a high inference accuracy with a small rectifier inside TEE while protecting critical model parameters and edge data. 
\item We implement a real-world deployment using Intel SGX to run GNNs locally. Experimental results show that the introduction of the enclave increases a small inference overhead and negligible accuracy degradation (less than $2\%$), demonstrating the effectiveness of \mymethod. 
\item We conduct a security analysis with link stealing attacks on the intermediate data communicated from the public backbone to the rectifier, showing that no private edge information is leaked in the untrusted environment.

\end{enumerate}

\section{Background} \label{sec: background}
\subsection{Graph Neural Network} \label{bg: gnn}
GNNs process graph-structured data, which contains relational information between nodes (or edges).
Without loss of generality, we follow the common GNN definition ~\cite{xu2018powerful}:
a graph $G=(V, E)$ consists of the node set $V$ with $n$ nodes, each with a feature represented by a $d$-dimensional vector, denoted as $X\in \mathbb{R}^{n\times d}$.
The set of edges $E$ is depicted by an adjacency matrix $A\in \{0, 1\}^{n\times n}$, indicating the existence of connections between nodes in $G$.
A GNN model is a function $f(X, A)$ to summarize global information in the graph for downstream tasks including node classification~\cite{zhou2019meta}, community detection~\cite{shchur2019overlapping}, and recommender systems~\cite{wu2019session}.
The forward propagation of GNN layer $k$ can be written as 

{\footnotesize
\begin{equation}
    H^{(k)} = \sigma(\hat{A}H^{(k-1)}W^{(k)})
\end{equation}}where $H^{(k)}\in \mathbb{R}^{N\times d^{(k)}}$ is the matrix of node embedding at layer $k$, $\hat{A}$ is the adjacency matrix with self-loops normalized with degree matrix, $W^{(k)}\in \mathbb{R}^{d^{(k-1)}\times d^{(k)}}$ is the learnable weight matrix, and $\sigma$ is an activation function. 
Note that GNN operations require $\hat{A}$ through message passing during the inference process, which causes edge data privacy concerns. 

\subsection{Model Deployment in Trusted Execution Environments} \label{bg: deployment}
A TEE, such as Intel SGX or ARM TrustZone, protects code and data within it, ensuring their confidentiality and integrity with hardware support against strong adversaries including a malicious host operating system.
The increasing demand for edge AI applications, e.g., those that require real-time responses or rely on local data or private data, has led to a trend of using TEEs for IP protection and privacy preservation.
Previous works focus on securely deploying DNNs by isolating the non-linear layers~\cite{sun2023shadownet} or by partitioning the model and placing the final dense layers inside the TEE \cite{mo2020darknetz}.
Recently, Zhang et.al~\cite{zhang2024no} also consider the privacy of training data and propose a ``partition-before-training'' strategy against membership inference attack~\cite{shokri2017membership}. 
Additionally, Liu et al.~\cite{liu2023mirrornet} design a deployment strategy that only allows single-directional data flow from the untrusted environment to the enclave. 
However, all previous works only focus on local DNN's IP and data privacy, which are not applicable to the secure deployment of a GNN on the edge. 
Our proposed method, \mymethod, bridges this gap by protecting the security of trained GNNs and private edge data using TEEs.

\section{Problem Statement} \label{sec: statement}

\subsection{Deploying GNN locally Causes Vulnerabilities} \label{ps: gnn inference}
Deploying GNNs on local devices requires access to graph data in addition to the trained model, which introduces unique security and privacy challenges. 
Similar to DNN deployment, the IP of the well-performed local model, including its trained weights and biases, is valuable asset that must be protected against model extraction attacks.
Beyond the model IP, local GNN inference raises additional privacy concerns due to the nature of GNN architecture. 
Specifically, during the message-passing phase of GNN inference, target nodes aggregate information from neighboring nodes to update their embeddings. 
This process involves accessing sensitive edge data, such as user-product interactions in recommender systems.
In our work, we will address the GNN IP infringement and edge data breach vulnerability during GNN deployment.

\subsection{Edge Privacy is Valuable} \label{motivation: edge importance}

\begin{figure}[t]
    \centering
    \includegraphics[width=0.95\linewidth]{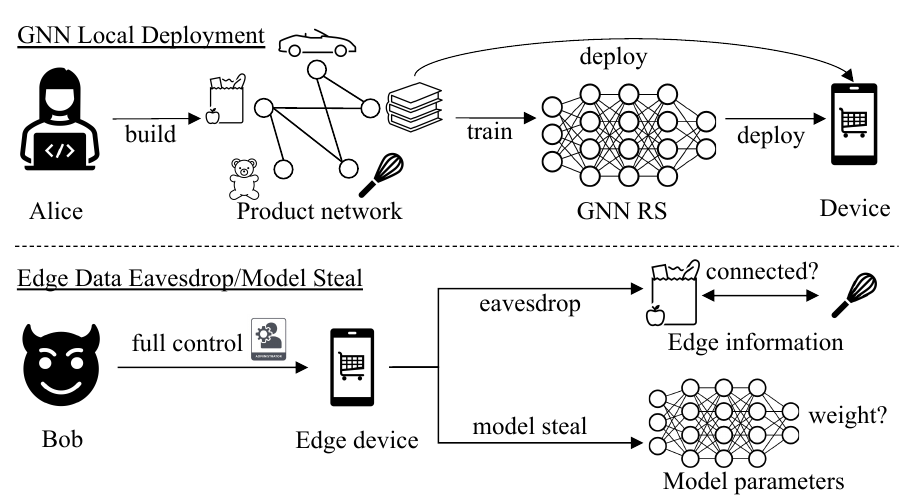}
    \caption{\textbf{Motivation Example:} Alice (victim) builds a graph of products and trains a GNN RS. She deploys both edge data and RS on local devices. Bob (attacker) accesses this device and steals the edge data and model parameters.}
    \label{fig: problem-statement}
\end{figure}

Membership inference attack is the most common data privacy threat to machine learning models~\cite{shokri2017membership}, where the goal is to determine whether a given data point belongs to the training set. 
However, in the context of GNNs, edge data raises additional privacy concerns. 
Link stealing attacks~\cite{he2021stealing, ding2023vertexserum} aim to infer the connectivity between any pair of given nodes. 
In this work, we focus on the adjacency information (edges), while considering the node features as public.
A real-world example is illustrated in Fig.~\ref{fig: problem-statement}, where Alice (victim) deploys a recommender system (RS) on local edge devices. 
In such a product graph, the node features are public attributes of the products—such as price, user reviews, or categories—that are available to any user. 
However, the internal relationships between products require intensive learning from user behavior data, which is valuable IP for the model vendor. 
Therefore, safeguarding the node connectivity information during GNN local inference is of great importance.

\subsection{TEE Has Memory Restrictions} \label{ps: TEE}
The introduction of TEE greatly enhances data security and privacy with secure compartments. 
However, TEE platforms face significant memory limitations, a critical constraint for secure computation. 
For instance, for Intel SGX trusted enclaves, the physical reserved memory (PRM) is limited to 128MB, with 96 MB of it allocated to the Enclave Page Cache (EPC)~\cite{intel2017sgx}. 
Excessive memory allocation will lead to frequent page swapping between the unprotected main memory and the protected enclave, which can cause high overhead and additional encryption/decryption to ensure data integrity~\cite{costan2016intel}.
This memory constraint poses a significant challenge for deploying GNN models and the entire graph (including node features and adjacency information) within the secure enclave, which often far exceed the PRM limitation of enclaves.

\begin{figure*}[t]
    \centering
    \includegraphics[width=0.99\linewidth]{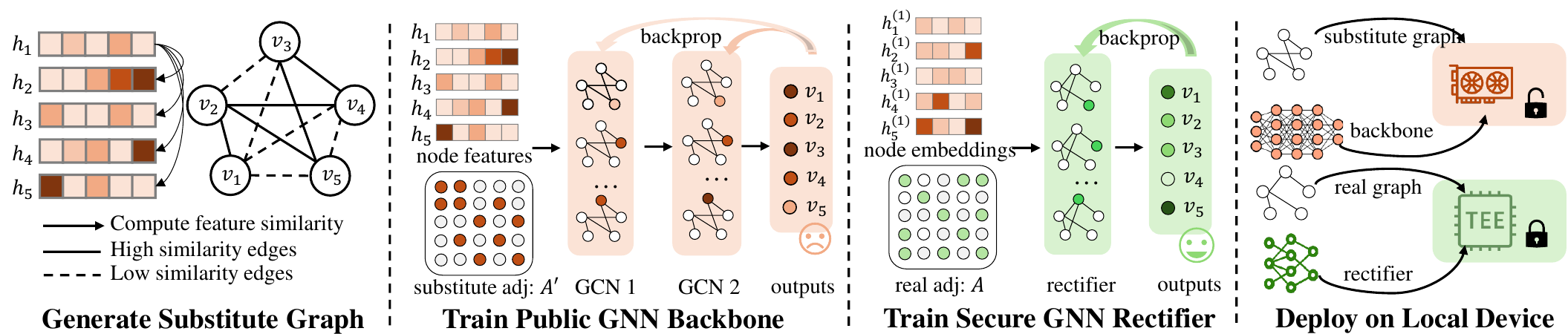}
    \caption{\textbf{Overview of \mymethod.} \mymethod can be summarized into four steps: 1. generate a substitute graph based on the similarity between the node features; 2. train the GNN backbone with the substitute adjacency matrix; 3. freeze the backbone and train the secure recalibration rectifier using real adjacency information; 4. deploy the backbone and substitute graph in untrusted environments (in {\color{brown} brown}) and deploy the real graph and rectifiers on enclaves (in {\color{green!30} green}). }
    \label{fig: overview}
    \vspace{-2mm}
\end{figure*}

\section{Our Approach: \mymethod } \label{sec: method}

\subsection{Threat Model} \label{method: threat model}
Our work aims to protect both GNN model parameters and the edge data during inference, specifically using the TEE of Intel SGX. 
Unless specified, we consider node classification as the primary downstream task.

\noindent\textbf{Attacker's Goal.}
We assume there are ``honest-but-curious'' attackers such as users of local ML systems.
Specifically, these passive attackers are curious about sensitive data but do not intend to corrupt the system’s integrity and functionality. 
Their objective is to extract confidential graph structural information, represented by the adjacency matrix, and to steal the parameters of highly accurate GNNs present on the device.

\noindent\textbf{Attacker's Capability.}
The attacker has full control over the victim system, enabling them to query the GNN model with any chosen node to obtain the model output results. 
Additionally, they can observe any computation and intermediate results in the untrusted environment, such as the public part of the GNN model or the intermediate node embeddings.
However, any data and operations within the enclaves remain sealed and inaccessible to the attacker. 
Other attacks against TEE, e.g., side-channel attacks, are out of the scope of this work.

\noindent\textbf{Attacker's Knowledge.}
We assume the attacker lacks knowledge of the adjacency matrix, which represents confidential relationships between any two nodes in the graph (e.g., hidden correlations between customers and products in recommender systems or private connections between users in social networks). 
The node features, however, are considered public knowledge such as product attributes or user profiles.

\noindent\textbf{Defender's Goal.}
The defender, a model provider such as Apple or Amazon, is responsible for the design and deployment of the GNN model on the local device, aiming to protect both the highly accurate GNN model and the private adjacency matrix information. This involves designing the GNN structure and establishing the secure deployment of the trained model and graph data on the inference platform.

\subsection{Overview of \mymethod} \label{method: overview}
\mymethod employs the design strategy known as \textit{partition-before-training}~\cite{zhang2024no}, to completely avoid using the private edge data during training of the public backbone. 
To maintain the model performance, a GNN rectifier is designed to use the real adjacency matrix, which is lightweight and securely protected within the TEE. 
Furthermore, to prevent information leakage through intermediate data transfer, \mymethod allows only one-way communication from the untrusted environment to the enclave. 
As illustrated in Fig.~\ref{fig: overview}, \mymethod can be summarized in four steps: (1) generate substitute graph(s); (2) train the public backbone; (3) train the private rectifier;  (4) securely deploy the data and models.

\subsection{Public GNN Backbone} \label{method: backbone}
The backbone is trained with public data, aiming to provide sufficient pre-computation during the inference process. Unlike the backbones for DNNs or LLMs in~\cite{zhang2024no}, which are open-source pre-trained models, GNNs are more dataset-specific and cannot directly leverage such models. Therefore, our \mymethod uses a pre-trained neural network backbone that utilizes public node features. We generate a substitute adjacency matrix based on node similarity and train this backbone for the target tasks (node classification).

\noindent\textbf{Substitute Adjacency Matrix:}
To enable the backbone model to capture message passing between nodes, we design it as a GNN with a substitute adjacency matrix. This substitute adjacency is computed based on the similarity between nodes:

{\footnotesize
\begin{equation}
    A'(i, j) = 
\begin{cases}
1 & \text{if } F(i, j)\geq\tau \text{ for each } j \neq i \\
0 & \text{otherwise}
\end{cases}
\end{equation}}Here, $F(i, j) = \text{sim}(x_i, x_j)$ represents the similarity measure computed between two node features, such as cosine similarity $\frac{x_i \cdot x_j}{|x_i||x_j|}$. Based on this similarity, we determine the connected edges using a threshold $\tau$. Alternatively, connectivity can be computed using the $k$-nearest neighbors, where we connect the top-$k$ similar nodes to construct the adjacency matrix.

\noindent\textbf{Training the Public Backbone:}
The public backbone consists of GCN layers trained with the substitute adjacency matrix $A'$ and the public node features $X$. Each GCN layer $k$ projects the high-dimensional features (embeddings) from $\mathbb{R}^{N \times d^{(k)}}$ to a lower-dimensional space $\mathbb{R}^{N \times d^{(k+1)}}$ through the message passing procedure. However, due to the loss of spatial information from the real adjacency matrix, this public backbone often exhibits lower accuracy. Therefore, the other component, a private rectifier, aims to recalibrate the embeddings in $\mathbb{R}^{N \times d^{(k+1)}}$ with real adjacency to achieve higher accuracy.

\subsection{Secure GNN Rectifier} \label{method: rectifier}

\begin{figure}[t]
    \centering
    \includegraphics[width=0.93\linewidth]{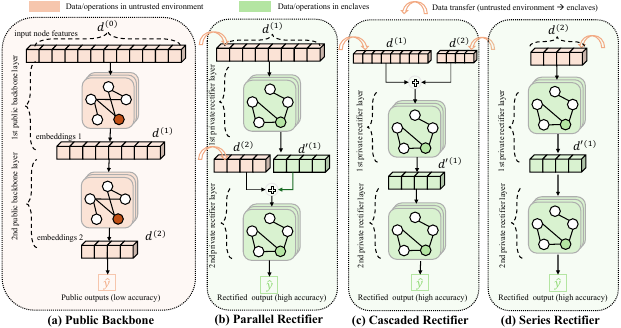}
    \caption{\textbf{Backbone and GNN Rectifier Designs:} (a). Public GNN backbone; (b). Parallel rectifier; (c). Cascaded rectifier; (d) Series rectifier.}
    \label{fig: designs}
    \vspace{-3mm}
\end{figure}

With the public backbone trained using the substitute adjacency matrix, the node embeddings
are passed from the backbone to the GNN rectifier to rectify the misinformation from $A'$.
The rectifier is designed with GCN layers that take the outputs of the backbone as inputs, along with the true adjacency matrix for rectification. 
Notably, the embeddings are often in a lower-dimensional latent space compared to the original node features. 
Therefore, a rectifier layer will be smaller than the corresponding backbone layer, so that it can fit into the enclave. 
During the training process of the rectifier, we freeze the pre-trained GNN backbone and adjust the rectifier parameters using cross-entropy loss for node classification.

How to process node embeddings from the backbone will result in various rectifier sizes and rectification performances. We consider three communication schemes (data flows) and the corresponding rectifiers, as shown in 
Fig.~\ref{fig: designs}:

\noindent\textbf{Parallel Rectifier:} Layers of the public backbone and layers of the rectifier run layer-by-layer in parallel. We aim to rectify the node embeddings right after each message-passing with the substitute adjacency matrix. Our experiment shows that the parallel rectifiers have the best accuracy (Sec.~\ref{exp: performance}). 
\ad{Generally, I would consider `performance' include both classifier accuracy and cost (time/memory). This here seems to only meaning accuracy. Should we be more specific in using the term?  }
\rd{revised.}

\noindent\textbf{Cascaded Rectifier:} This design allows the backbone to run first, and then we concatenate all GCN layer outputs from the untrusted environment as input to the rectifier, which provides the rectifier a global view of the node embeddings but introduces additional computational requirements.

\noindent\textbf{Series Rectifier:} In this design, we utilize only the final layer from the backbone as input to the rectifier. This approach uses the smallest input space and results in lower memory consumption and faster response time.

\subsection{Deployment of GNN on TEE} \label{method: deployment}
As shown in Fig.~\ref{fig: overview}, we deploy the public backbone and substitute graph in the untrusted environment, where the model can benefit from acceleration from GPUs.
We safeguard the real adjacency information and GNN rectifiers inside the secure enclaves to prevent edge privacy breaches and model stealing attacks.
The true adjacency matrix will be saved in the Coordinate (COO) format, which maintains only the non-zero elements with their indices for memory efficiency, with the pre-computed degree matrix, to accelerate the normalization process.
During the inference phase, each layer of the rectifier's output will be stored internally within the enclave, preventing intermediate data from leaking to the untrusted environment.
Logits have been shown to contain additional privacy information such as data membership and link connectivity~\cite{he2021stealing, shokri2017membership}.
Therefore, we also require the final output to be label-only, i.e., the logits are kept inside the enclave and only the output class is available to users.

\section{Experimental  Validation} \label{sec: experiment}

\subsection{Experiment Setup} \label{exp: setup}

\noindent\textbf{Datasets:} 
\begin{table}[h]
    \centering
    \vspace{-5mm}
    \caption{Dataset Used in \mymethod Validation}
    \label{tab: dataset}
    \resizebox{\linewidth}{!}{
    \begin{tabular}{l|c|c|c|c|c}
    \hline
    \rowcolor{lightgray}
    Dataset  &  $\#$ Node  & $\#$ Edge & $\#$ Feature & $\#$ Class  & Dense $A$ (MB) \\
    \hline
    Cora~\cite{yang2016revisiting}     &  2,708  &  10,556  & 1,433 & 7 & 167.85 \\
    \hline 
    Citeseer~\cite{yang2016revisiting}     &  3,327  &  9,104  & 3,703 & 6 & 253.35 \\
    \hline
    Pubmed~\cite{yang2016revisiting}  & 19,717 & 88,648 & 500 & 3 & 8898.01\\
    \hline
    Computer~\cite{shchur2018pitfalls}    & 13,752  &491,722  &767 &  10 & 4328.56 \\
    \hline
    Photo~\cite{shchur2018pitfalls} &7,650  & 238,162  & 745  & 8 & 1339.47 \\
    \hline
    CoraFull~\cite{bojchevski2017deep}  &  19,793 &126,842  &8,710 & 70 & 8966.74 \\
    \hline
    \end{tabular}
    }
\end{table}
We focus on semi-supervised learning tasks for node classification with GNNs on multiple datasets at different scales, as shown in Table~\ref{tab: dataset}.
Specifically, we evaluate \mymethod on three standard datasets in~\cite{yang2016revisiting}, two larger graphs \textit{Amazon Computer} and \textit{Photo} in~\cite{shchur2018pitfalls}, and \textit{CoraFull}~\cite{bojchevski2017deep} which has many classes ($70$).
Table~\ref{tab: dataset} shows the memory requirement for a dense adjacency matrix.
To train the GNN, we follow the common practice: using $20$ labeled node per class~\cite{shchur2018pitfalls}, and the unlabeled nodes are used as the testing set.
If not specified, we use KNN with $k$=$2$ as substitute graphs in backbones, which are chosen based on the ablation study in~\ref{exp: ablation}.

\yf{here rather than giving the specific channel size, it is better to present design rationale, such as the input channel size is determined by the node feature dimension? is the output channel size determined by anything or you just freely choose it?}\rd{revised}
\noindent\textbf{Models:}
We evaluate \mymethod on three different GNN structures based on dataset features space size:
$\mathcal{M}_1$ has as a 3-layer GCN backbone with output channel size as $(128, 32, \mathcal{C})$ and a rectifier $(128, 32, \mathcal{C})$, which is designed for a smaller dataset including Cora, Citeseer, Pubmed, where $\mathcal{C}$ is the label space size.
$\mathcal{M}_2$ has wider output channels ($256$) for both the backbone and the rectifier, which is used for the case when the model has a high-class number, such as CoraFull.
We also test a larger and deeper design $\mathcal{M}_3$ with a backbone $(256, 64, 32, 16, \mathcal{C})$ and rectifier $(64, 32, \mathcal{C})$. 
Different rectifiers may have various input channel sizes, as shown in Fig.~\ref{fig: designs}.

\noindent\textbf{Metrics:}
We first compute the accuracy of the original GNN model, which has the same architecture as the backbone but uses the real adjacency matrix and therefore has different parameters, denoted as $p_{org}$.
The difference between the classification accuracy of the public backbone in the normal world, denoted as $p_{bb}$, and the accuracy of the GNN rectifier, denoted as $p_{rec}$, measures the \textit{protection performance}. 
We also compare the size of the public backbone and private rectifier in terms of the number of model parameters, denoted as $\theta_{bb}$, and $\theta_{rec}$.
To show accuracy degradation, we utilize the difference between $p_{rec}$ and $p_{org}$ and the silhouette score to show the node's clustering performance~\cite{shahapure2020cluster}.

\noindent\textbf{Platform:}
We run the training procedure and the software evaluation process on Ubuntu 18.04.6 with NVIDIA TITAN RTX.
We also follow a similar setting to~\cite{zhang2024no} as a real-world case study on an SGX-enabled Intel Core i7-7700, running Ubuntu 20.04.6 and utilizing Intel SGX SDK and PSW.

\subsection{Software-based Validation} \label{exp: software}
\begin{table*}[t]
    \centering
        \caption{\mymethod Performance with KNN  graph ($k=2$), \textbf{Bold indicates the best rectifier}; 
    \underline{Underline is the second best}.}
    \label{tab: performance}
    \resizebox{\linewidth}{!}{
    \begin{tabular}{l||c|c|c||c|c|c||c|c|c||c|c|c}
    \hline
    \rowcolor{lightgray}
    & \multicolumn{3}{c||}{Original / Backbone}  & \multicolumn{3}{c||}{Parallel \mymethod}  & \multicolumn{3}{c||}{Series \mymethod}  & \multicolumn{3}{c}{Cascaded \mymethod}\\
    \hline
    Dataset       & $p_{org}$(\%) & $\theta_{bb}$ (M) & $p_{bb}$ (\%) & $p_{rec}$ (\%)& $\Delta p$ (\%)& $\theta_{rec}$ (M)& $p_{rec}$ (\%) &  $\Delta p$ (\%) & $\theta_{rec}$ (M) & $p_{rec}$ (\%) & $\Delta p$ (\%) & $\theta_{rec}$ (M)\\
    \hline
    \hline
    Cora~\cite{yang2016revisiting}  &   80.4 &  0.188 &  60.2   &   \textbf{78.8}     &  18.6  &  0.022        &   \underline{78.2}    &   18.0    &    0.0088      &   77.6    &    17.4    &    0.027      \\
    Citeseer~\cite{yang2016revisiting}  & 65.2 & 0.479 &  60.3   &   \textbf{70.1}    &  9.80  &  0.022        &    68.7   &    8.40    &    0.0087      &   \underline{69.0}    &    8.70    &    0.026      \\
    Pubmed~\cite{yang2016revisiting}  &   77.1 & 0.068  &   66.6    &   \textbf{75.2}    &  8.60  &  0.022         &   \underline{75.1}    &    8.50    &    0.0085      &   73.6    &    7.00   &    0.025      \\
    Computer~\cite{shchur2018pitfalls}  &  75.5  & 0.216 &   56.6   &   \underline{77.6}     &  21.0 &  0.021        &   \textbf{78.2}   &    21.6    &    0.0039     &   77.4    &    20.8   &    0.027      \\
    Photo~\cite{shchur2018pitfalls}  &  83.7   & 0.210 &  68.3    &   \underline{84.9}     &  16.6 &  0.021            &   84.2    &   15.9    &    0.0037     &   \textbf{85.1}   &   16.8    &    0.026     \\
    CoraFull~\cite{bojchevski2017deep}  &   59.5 & 2.27  &  43.1    &   \underline{57.8}     &  14.7  &  0.051          &   \textbf{58.0}    &   14.9     &    0.050     &   55.8    &    12.7    &    0.060      \\
    
    \hline
    \end{tabular}
}
\end{table*}
\begin{table}[t]
\vspace{-3mm}
    \centering
    \caption{Compare various backbone designs}
    \resizebox{\linewidth}{!}{
    \begin{tabular}{l||c|c||c|c||c|c||c|c}
    \hline
    \rowcolor{lightgray}
    &     \multicolumn{2}{c||}{DNN} & \multicolumn{2}{c||}{random} & \multicolumn{2}{c||}{cosine} & \multicolumn{2}{c}{KNN} \\
    \hline
    Dataset     & $p_{bb}$ & $p_{rec}$ & $p_{bb}$ & $p_{rec}$ & $p_{bb}$ & $p_{rec}$ & $p_{bb}$ & $p_{rec}$\\
    \hline
    Cora    & {54.4} & 76.8 & 17.2 & 51.5 & 55.3 & \textbf{79.1} & 60.2 & \underline{78.8} \\
    Citeseer & 53.9 & \underline{64.6} & 18.9 & 38.3 & {46.2}  & 64.3 & 66.6 & \textbf{70.1} \\
    Pubmed    & 71.9 & 73.9 & 34.5 & 52.1 & 72.1 & \textbf{76.0} & {66.6} & \underline{75.2}\\
    Computer & 52.6 & 73.6 & 7.16 & 28.9 & {44.6} & \underline{76.7} &56.6 & \textbf{77.6} \\
    Photo & {64.3} & 83.4 & 30.4 & 52.8 & 69.1 & \textbf{84.9} & 68.3 & \underline{84.9}\\
    CoraFull & 43.9 & \underline{57.7} & 2.69 & 27.3 & {40.1} & 55.6 & 43.1 & \textbf{57.8} \\
    \hline
     \end{tabular}
}
\vspace{-2mm}
    \label{tab: backbone}
\end{table}

\subsubsection{Evaluation of \mymethod Performance}
\label{exp: performance}
The performance of GNN deployment using TEE on a local device can be evaluated using the following attributes: 1. Accuracy degradation ($p_{\text{org}}$ - $p_{\text{rec}}$) (lower is better); 2. Protection performance $\Delta p = p_{\text{rec}} - p_{\text{bb}}$ (higher is better); 3. The model size inside the enclave $\theta_{rec}$ (smaller is better).
In Table~\ref{tab: performance}, we present the performance of \mymethod on three rectifier designs, following the experimental setups described in Section~\ref{exp: setup}. Among the three rectifier designs, the parallel rectifier achieves %
the highest accuracy on three datasets and ranks second-best in others, and it also offers the highest protection. 
The cascaded rectifier has the largest model size but shows lower accuracy, as the input GCN layer may not handle all complex inputs together well.
The series rectifier has the smallest size inside the TEE, as it only requires the final node embedding from the backbone as input, and yet it still achieves good performance overall. 

\subsubsection{Evaluation of Different Backbones}
\label{exp: backbones}
\mymethod's backbones show low accuracy when using the substitute graph, indicating that our design effectively prevents malicious attackers from accessing any high-precision models in untrusted environments.
We then compare the performance of \mymethod using different types of backbones: the DNN backbone (an MLP using only node features), and the GNN backbones with three different ways to generate the substitute graph: random  (the graph is randomly generated), cosine (the graph is generated based on cosine similarity), and KNN (the graph is generated using the KNN algorithm). For the GNN-based backbones, we sample the graph density to match that of the original graph. Among all four types of backbones, the cosine and KNN methods demonstrate the best performance, while the GNN with a random graph shows the lowest accuracy. Specifically, the random graph incorporates too much misinformation, degrading the embedding performance of the backbone model and making rectification difficult. Moreover, compared with the DNN model, the GNN backbones using cosine similarity and KNN substitute graphs enable the backbone not only to extract information from the node features but to gather additional information from similar nodes via message passing. This additional information enhances the rectification process when using the correct adjacency information.

\subsubsection{Rectifier Interpretation}

\begin{figure}[t]
\vspace{-3mm}
    \centering
    \includegraphics[width=\linewidth]{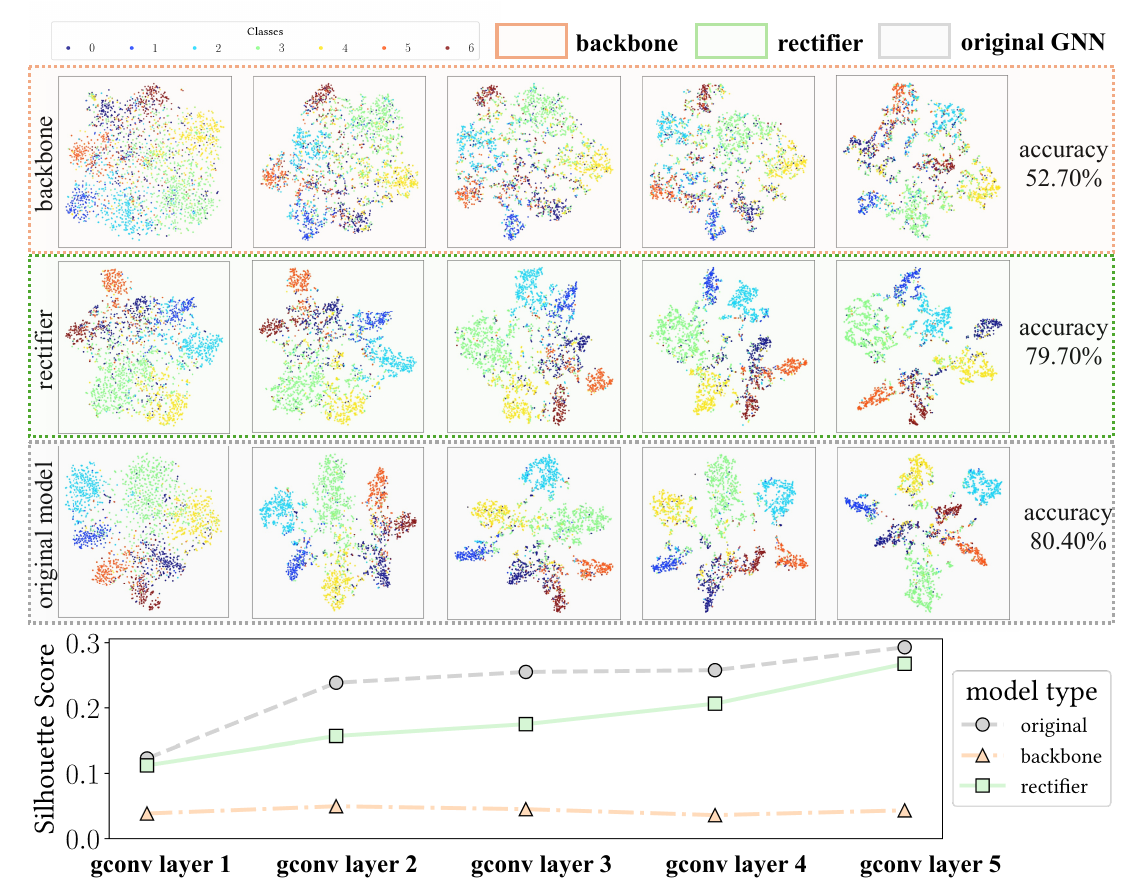}
    \caption{\textbf{Visualization of the node embeddings.} We visualize the embedding latent space for parallel \mymethod on Cora dataset using the same structure for the original GNN, backbone and rectifier as \textit{$\mathcal{M}_2$}. The line chart shows the Silhouette Score of the clustering performance\textbf{(the higher the better).}}
    \label{fig: visualizations}
    \vspace{-2mm}
\end{figure}

To illustrate the rectification process, we visualize the node latent space of a parallel \mymethod applied to the Cora dataset, which contains 7 classes. In Fig.~\ref{fig: visualizations}, we illustrate how the latent space changes layer-by-layer between the backbone and the rectifier using t-SNE~\cite{van2008visualizing}. The rectifier layer significantly improves the clustering performance by gradually incorporating the real adjacency matrix, ultimately approaching the original model's performance. Additionally, we plot the silhouette score of each layer's embeddings to numerically assess their clustering performance. The results show that the rectifier's scores (\textcolor{green!40}{green} line) become close to those of the original GNN (\textcolor{gray}{gray} line), while the backbone model's scores (\textcolor{orange!40}{orange} line) remain low.

\subsubsection{Ablation Study}\label{exp: ablation}
\begin{figure}[t]
    \centering
    \includegraphics[width=0.99\linewidth]{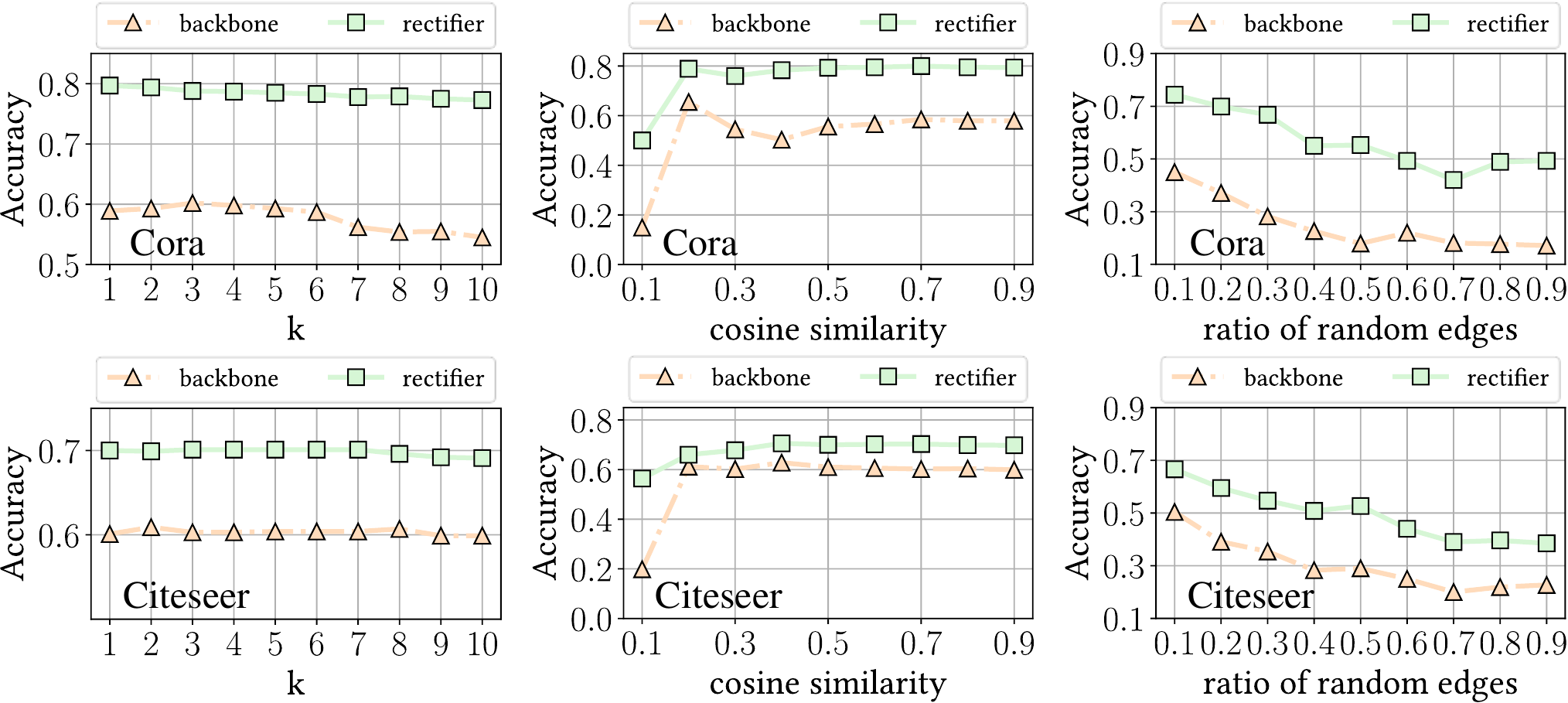}
    \caption{Impact of hyperparameters of different substitute graphs}
    \label{fig: ablations}
    \vspace{-2mm}
\end{figure}

We evaluate the impact of different hyperparameters on \mymethod's performance, as shown in Fig.~\ref{fig: ablations}. Specifically, we vary the number of neighbors in the KNN-based substitute graph, adjust the cosine similarity threshold for the cosine similarity graph, and change the percentage of random edges (relative to the number of real edges) for the random substitute graph.
For the KNN method, the performance remains stable because the algorithm consistently connects similar nodes based on their features. Increasing $k$ primarily affects the density of the substitute graph without significantly impacting performance. We select $k$=$2$ as the number of edges for it is close to the real graph for most of the datasets. In contrast, selecting a low cosine similarity threshold for the cosine similarity graph can connect unrelated nodes, which adversely affects \mymethod's performance (when the cosine similarity threshold is $\leq 0.2$). As we increase the number of random edges in the substitute graph, the performance of both the backbone and the rectifier decreases as more structural noise is introduced. Notably, when we use an extremely small number of edges, random \mymethod's accuracy approaches that of the DNN backbone.

\subsection{Real-world Implementation} \label{exp: implementation}

\subsubsection{Implementation Details} 
\rd{talk about the SGX implementation of SGX}
We implemented GNNVault on a desktop PC equipped with an Intel Core i7-7700 (3.60GHz) processor. The public backbone in the untrusted environment is developed using PyTorch 2.4.1, while the rectifier is implemented in C++ with Intel SGX SDK 2.25. 
To optimize the computation within the rectifier, we leveraged the Eigen linear algebra library to efficiently handle matrix operations, achieving approximately 20\% speed up for large models~\cite{eigen}. 
\yf{rephrase the last line}\rd{I remove the last line, I think no need to mention what implementation our code is based on.} 

\subsubsection{Inference Overhead and Memory Usage}
\begin{figure}[t]
    \centering
    \includegraphics[width=\linewidth]{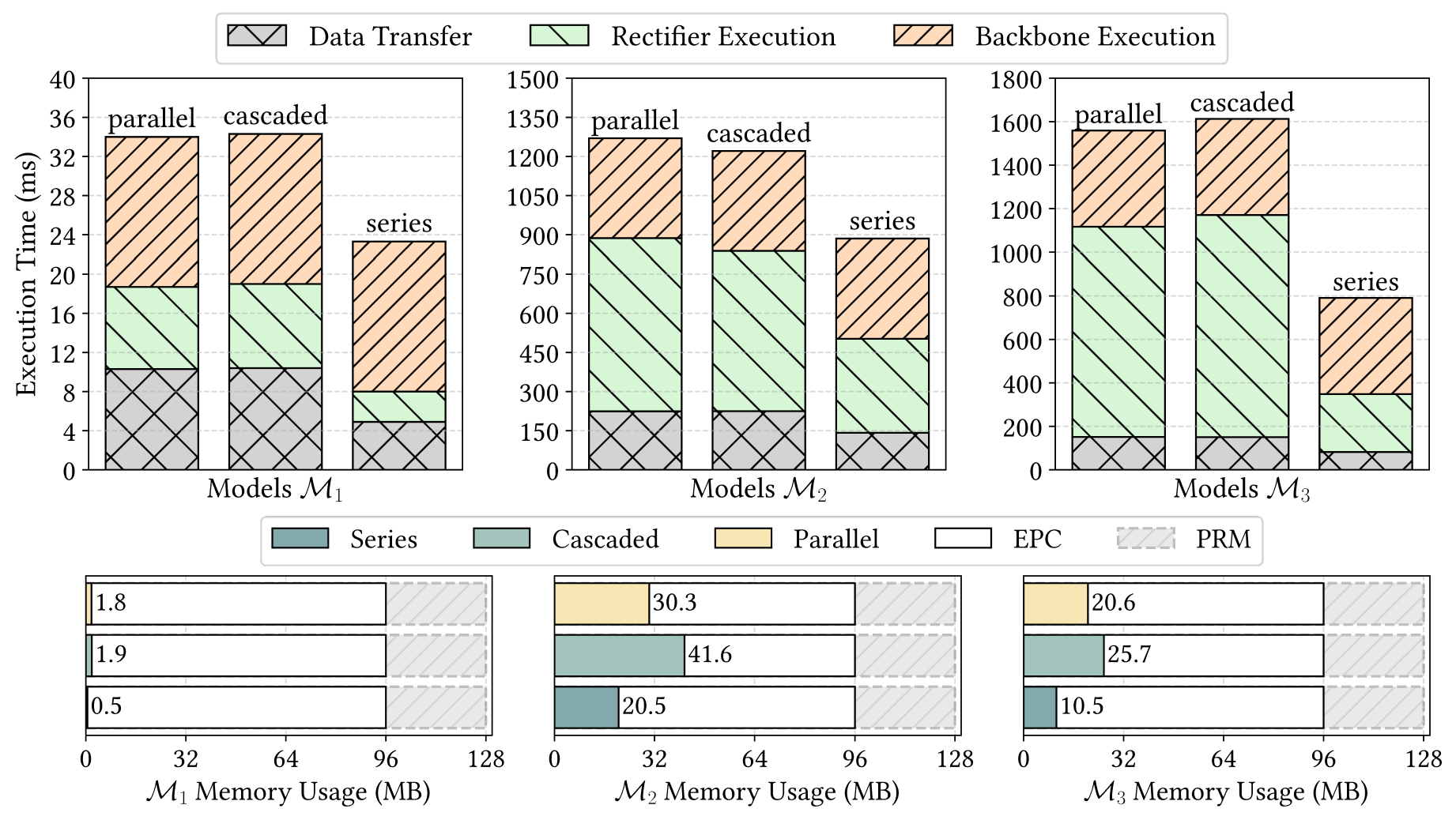}
    \caption{\mymethod Inference Time Breakdown and Memory Usage}
    \label{fig: execution time}

\end{figure}

The inference time of \mymethod comprises of the backbone execution time, data transfer time, and the rectifier execution time. 
In Fig.~\ref{fig: execution time} (top), we present a breakdown of the inference time, for three rectifier designs and different GNN models: $\mathcal{M}_1$ (Cora), $\mathcal{M}_2$ (CoraFull), and $\mathcal{M}_3$ (Amazon Computer). The parallel and cascaded rectifiers transfer all intermediate node embeddings from the normal world to the secure world, resulting in higher latency due to data transfer and execution inside the TEE. The series rectifier exhibits the smallest overhead since it only requires the last layer's embedding, leading to an overhead of approximately $52\%$ to $131\%$ compared with running an unprotected GNN with CPU. Note that the backbone can be accelerated using GPUs to further increase the inference speed.

\rd{talk about the memory usage, and point out if it applicable to put the entire GNN inside the TEE.}

We also present the enclave runtime memory usage required for different GNNVault configurations, as shown in Fig.~\ref{fig: execution time} (bottom). As mentioned before, the memory allocated in the trusted enclave is limited by SGX's PRM and EPC.
Therefore, it is essential to %
ensure that the rectifier operates effectively within the enclave's secure boundaries and does not exceed the recommended limits.
The enclave memory usage is primarily for each layer’s input features, adjacency matrix, and model parameters. Notably, the highest memory usage across all models and rectifier designs remains at only 41.6MB, well within the 96MB EPC threshold. This confirms the feasibility of running these configurations securely within the enclave. In contrast, the backbone’s runtime memory usage for the three models significantly exceeds 128MB, underscoring the impracticality of executing a complete GNN model within the enclave due to memory constraints.

\subsection{Security Analysis using Link Stealing Attacks}
\begin{table}[ht]
    \vspace{-3mm}
    \centering
    \caption{Link Stealing Attack Performance on \mymethod}
    \resizebox{\linewidth}{!}{
    \begin{tabular}{l||c|c|c||c|c|c||c|c|c}
    \hline
    \rowcolor{lightgray}
      
     Dataset &  $\mathcal{M}_{org}$   & $\mathcal{M}_{gv}$  &  $\mathcal{M}_{base}$ &  $\mathcal{M}_{org}$   & $\mathcal{M}_{gv}$  &  $\mathcal{M}_{base}$ &  $\mathcal{M}_{org}$   & $\mathcal{M}_{gv}$  &  $\mathcal{M}_{base}$ \\
    \hline 
    & \multicolumn{3}{c||}{Euclidean}  & \multicolumn{3}{c||}{Correlation} & \multicolumn{3}{c}{Cosine} \\
    \hline 
     Cora  & 0.844  &  \textbf{0.702} & 0.715 & 0.903 &\textbf{0.735} & 0.720 & 0.972 & \textbf{0.765 } & 0.754  \\
    Citeseer  & 0.915  &  \textbf{0.750} & 0.731 & 0.912 & \textbf{0.778} & 0.752 & 0.987 & \textbf{0.807} & 0.790 \\
    \hline
    & \multicolumn{3}{c||}{Chebyshev}  & \multicolumn{3}{c||}{Braycurtis} & \multicolumn{3}{c}{Canberra} \\
    \hline 
     Cora  & 0.847  &  \textbf{0.661} & 0.691 & 0.902 &\textbf{0.696} & 0.693 & 0.933 & \textbf{0.741} & 0.717  \\
    Citeseer  & 0.908  &  \textbf{0.711} & 0.698 & 0.953 & \textbf{0.751} & 0.732 & 0.976 & \textbf{0.785} & 0.746 \\

    \hline
    \end{tabular}
    }
    \label{tab: sla attack}
\end{table}

Link stealing attacks~\cite{he2021stealing, ding2023vertexserum} aim to infer the adjacency matrix by exploiting the similarity between node embeddings, based on the assumption that GNN layers produce more similar embeddings for connected nodes than for unconnected ones. \mymethod isolates the private graph within the secure world and ensures one-directional communication from the normal world to the secure world. To verify the effectiveness of this isolation, we conduct link stealing attacks using six different similarity metrics and present the attack ROC-AUC scores in Table~\ref{tab: sla attack} using all intermediate embeddings. A higher AUC score indicates greater leakage of private graph information. We use the attack performance on a DNN model with pure node features as the baseline ($\mathcal{M}_{\text{base}}$). The unprotected GNN ($\mathcal{M}_{org}$) shows very high AUC scores using all similarity metrics on both datasets, but with our \mymethod ($\mathcal{M}_{gv}$), the attack performance decreases to the level of the $\mathcal{M}_{base}$.
\yf{what does gv stand for?} \rd{revised}

\section{Conclusions and Future Work} \label{sec: conclusions}
In this paper, we propose a novel secure GNN deployment strategy called \mymethod. 
Our approach involves strict isolation,  training a public backbone model using only public data, and training a rectifier on private data to rectify the results from the backbone. %
The two spited models and their corresponding graph data are deployed in the untrusted environment and secure enclave, respectively. 
This method effectively safeguards sensitive adjacency information during GNN inference on edge devices. 
Given the growing trend of GNN local deployment, ensuring data privacy in such settings is of paramount importance. 
As part of our future work, we plan to explore and implement our strategy with additional GNN architectures, including GraphSAGE and GAT, to further enhance the security and broaden applicability.

\bibliographystyle{IEEEtran}
\bibliography{ref}

\end{document}